\documentclass[doublecol]{epl2}

\usepackage{amssymb,amsmath}
\usepackage{graphicx}
\usepackage{dcolumn}
\usepackage{bm}
\usepackage{color}
\usepackage{float}
\usepackage{hyperref}
\usepackage[version=4]{mhchem}
\usepackage{relsize}
\usepackage{comment}
\usepackage{media9}

\usepackage{booktabs}
\usepackage{algorithm}
\usepackage{algpseudocode}
\usepackage{siunitx}

\definecolor{RevTwo}{rgb}{0.00,0.00,0.00}
\definecolor{RevThree}{rgb}{0.00,0.00,0.00}
\definecolor{RevFour}{rgb}{0.00,0.00,0.00}
\definecolor{RevAll}{rgb}{0.00,0.00,0.00}

\newcommand{\RII}[1]{{\color{RevTwo}#1}}
\newcommand{\RIII}[1]{{\color{RevThree}#1}}
\newcommand{\RIV}[1]{{\color{RevFour}#1}}
\newcommand{\RM}[1]{{\color{RevAll}#1}}

\newcommand{\Ez}{{E_z}}
\newcommand{\Hx}{{H_x}}
\newcommand{\Hy}{{H_y}}
\newcommand{\CFL}{S}

\title{Ternary-Valued Finite-Difference Time-Domain Method: \RII{Asymptotic Agreement} with the Yee Scheme Through Noise-Shaped Quantisation}
\shorttitle{Ternary-Valued Finite-Difference Time-Domain Method} 

\author{Ivan S.~Maksymov}
\shortauthor{Ivan S.~Maksymov}

\institute{
  Seymour Research Laboratories, Seymour, VIC 3660, Australia
}

\abstract{I show that finite-difference time-domain (FDTD) dynamics can be reproduced using field variables restricted to the ternary alphabet ${-1,0,+1}$. The integration is carried out by a state accumulator, while quantisation is done by a second-order noise-shaped encoder with a separate error register. The Courant number $\CFL$ plays two roles: it is the exact fixed-point ratio and also sets the encoder's oversampling ratio. \RM{When $\CFL$ is chosen as a power of two, each cell update requires neither run-time multiplication nor floating-point arithmetic, and each emitted field symbol is represented by just two bits.} \RII{The ternary scheme converges asymptotically to the standard Yee scheme as $\CFL$ decreases towards zero.} \RIII{I report wall-clock and instruction counts relative to a floating-point reference, long-time energy integrations over $2\times10^{5}$ steps and the $\CFL$ required to achieve a prescribed accuracy for the driven mode.} \RII{Extending the approach to acoustics, Virieux-type elastodynamics and Schr{\"o}dinger-equation solvers is a prospective direction towards a broader class of quantised physics solvers for resource-constrained and specialised hardware.}}

\begin{document}

\maketitle


\section{Introduction}
The finite-difference time-domain (FDTD) method~\cite{Taf80, Zio83} is a key computational technique in physics and engineering, with applications ranging from electromagnetism and optics~\cite{Taflove, Sullivan, oskooi2010meep} to acoustics~\cite{Sullivan, Schneider_FDTD, botteldooren1994}, magnonics~\cite{Mak_FDTD}, elastodynamics~\cite{Vir84, Bossy2011}, quantum mechanics~\cite{Sullivan, Nag09, delgado2022relativistic} and sociophysics~\cite{Mak_FDTD}. The Yee algorithm~\cite{Yee66}, which underpins electromagnetic FDTD, discretises Maxwell's curl equations on a staggered grid and advances the electric and magnetic fields in a leapfrog sequence. Its computational cost per cell is dominated by the storage of multiple field components at finite numerical precision. These costs have motivated a long line of reduced-precision approaches, ranging from fixed-point implementations~\cite{Mak_FDTD} to specialised schemes~\cite{yu2013valu}, including field-programmable gate arrays (FPGAs)~\cite{Pfeifle2015, Mansoori}.

In parallel, and largely independently, similar challenges are being addressed in machine learning~\cite{Par25}. Transformer models such as BitNet~\cite{Ma24}, particularly in edge computing and FPGA implementations~\cite{Qia26}, employ reduced-precision representations and quantisation schemes to reduce memory requirements and computational cost. At the extreme, each neural weight can be represented by a single ternary digit, taking only the values $-1$, $0$ and $+1$~\cite{Ma24}. This representation is attractive for hardware implementation~\cite{Qia26}:~the spatial difference operators reduce to additions and comparisons, multipliers are eliminated from the update and data storage is reduced to two bits per neural weight.

I apply the same principle to the FDTD method, with the aim of reducing computational cost and enabling implementation on specialised hardware platforms. While the effects of reduced numerical precision in transformers are now well understood in machine learning~\cite{Ma24, Qia26}, a more fundamental question arises for FDTD:~how much of the underlying physics survives quantisation? A three-level field cannot represent an amplitude directly; it can instead encode one, much as a sigma-delta modulator represents an analogue waveform by an oversampled one-bit stream, with the signal encoded in the local density of output symbols rather than in their instantaneous values~\cite{Cal15}.

\RII{I also develop a comparison methodology appropriate for noise-shaped ternary fields. This is motivated by the fact that a ternary field compared naively with a floating-point reference can appear uncorrelated even when it correctly represents the underlying dynamics, because the comparison is made in a frequency band where the encoder has deliberately concentrated its quantisation error.}

Throughout, I work with the 2D transverse-magnetic system on a square Yee grid with $\Delta x=\Delta y$, unit wave speed and perfect electric conductor (PEC) walls, and I take the Courant number~\cite{Taflove}
\begin{equation}
  \CFL = \frac{c\,\Delta t}{\Delta x}\,
\end{equation}
as the single free parameter of the discretisation. The reference scheme is
\begin{align}
  \Hx|_{i,j} &\leftarrow \Hx|_{i,j} + \CFL\left(\Ez|_{i,j}-\Ez|_{i,j+1}\right)\,,
  \label{eq:yee-hx}\\
  \Hy|_{i,j} &\leftarrow \Hy|_{i,j} + \CFL\left(\Ez|_{i+1,j}-\Ez|_{i,j}\right)\,,
  \label{eq:yee-hy}\\
  \Ez|_{i,j} &\leftarrow \Ez|_{i,j} + \CFL\Big[\left(\Hy|_{i,j}-\Hy|_{i-1,j}\right)
  \nonumber\\
  &\qquad\qquad\quad -\left(\Hx|_{i,j}-\Hx|_{i,j-1}\right)\Big]\,,
  \label{eq:yee-ez}
\end{align}
with $\Ez$ forced to zero on the four boundaries and a soft source added to $\Ez$ at the grid centre. \RII{Throughout, Eqs.~\eqref{eq:yee-hx}--\eqref{eq:yee-ez} in single-precision arithmetic define the floating-point reference.}

\section{Separation of State and Encoder}
I introduce and distinguish the \emph{state}, which integrates the curl and must never be drained, from the \emph{encoder}, which converts the state into a ternary symbol and carries its own error register. Likewise, the threshold $T$ should not simultaneously determine the amplitude and the time step. These roles must be separated:~the quantisation determines how much field one symbol represents, while the Courant number determines how rapidly the state integrates.

I adopt the following fixed-point convention. A symbol $s\in\{-1,0,+1\}$ represents the physical field value $sA$, where $A$ is a full-scale amplitude chosen to exceed the largest field the simulation will produce. A state $a\in\mathbb{Z}$ represents the physical field value $aA/M$, where $M$ is the number of fixed-point units per symbol. The Courant number enters through the
integer
\begin{equation}
  M_C = M\,\CFL\,,
\end{equation}
which is required to be exact. With $M=10^5$ this admits
$\CFL=10^{-1},5\times10^{-2},2.5\times10^{-2},1.25\times10^{-2}$ and
$6.25\times10^{-3}$, which are the values used below. \RII{Alternatively, taking $M$ and $\CFL$ both as powers of two, for example $M=2^{17}$ with $\CFL=2^{-3},\ldots,2^{-7}$, is more restrictive but makes $M$ and $M_C$ powers of two simultaneously. Although the two sets do not correspond to exactly the same numerical $\CFL$ values, the second approach can be advantageous because the encoder's division and multiplication, as well as the clamp thresholds, can all be implemented as binary shifts, whereas a general choice of $M$ requires the compiler to implement the division using a multiply-high operation with a precomputed reciprocal constant.}

The state updates are then
\begin{align}
  a^{H_x}_{i,j} &\leftarrow a^{H_x}_{i,j} + M_C\left(s^{E}_{i,j}-s^{E}_{i,j+1}\right)\,,\\
  a^{H_y}_{i,j} &\leftarrow a^{H_y}_{i,j} + M_C\left(s^{E}_{i+1,j}-s^{E}_{i,j}\right)\,,\\
  a^{E}_{i,j}   &\leftarrow a^{E}_{i,j} + M_C\Big[\left(s^{H_y}_{i,j}-s^{H_y}_{i-1,j}\right)
  \nonumber\\
  &\qquad\qquad\quad -\left(s^{H_x}_{i,j}-s^{H_x}_{i,j-1}\right)\Big]\,.
  \label{eq:statefree}
\end{align}
The full-scale amplitude $A$ cancels identically from the update:~it appears only in the source injection and in the reconstruction. Every operation is an integer addition or a comparison.

\subsection{The encoder}
The state is converted to a symbol by a second-order error-feedback modulator
\begin{align}
  v &= x + \tfrac{1}{2}\left(3e_1 - e_2\right)\,,\\
  s &= \operatorname{clamp}\!\left(\left\lfloor v/M \right\rceil, -1, +1\right)\,,\\
  e_2 &\leftarrow e_1, \qquad e_1 \leftarrow v - sM\,,
\end{align}
where $x$ is the state, $\lfloor\cdot\rceil$ denotes rounding to nearest and $e_1,e_2$ are the encoder's own registers, one pair per field component per cell. The noise transfer function,
\begin{equation}
  N(z) = \left(1-z^{-1}\right)\left(1-\tfrac{1}{2}z^{-1}\right)\,,
  \label{eq:ntf}
\end{equation}
suppresses the quantisation-noise power at low frequency as $f^2$. \RII{In the simulations presented here, this choice keeps the encoder registers bounded while using only three quantiser levels, and makes the anti-windup clamp much less active than the canonical second-order choice. These results are empirical rather than guaranteed. The clamp, however, ensures that $e_1$ remains within $\pm2M$ at every call, so the encoder state is bounded by construction, regardless of the input. Bounded registers alone do not ensure an accurate field representation; that question is addressed separately by the convergence study in Table~\ref{tab:convergence}.}

\begin{table}[t]
\caption{\label{tab:convergence}Convergence at fixed physical time, $N\CFL=200$. The relative error is the Euclidean norm of the difference between the ternary \RII{state field $a^{E}A/M$} and floating-point $E_z$ fields, normalised by the floating-point norm over the full grid. The energy ratio compares the corresponding electric-field energies. \RII{The fitted convergence order is $p=0.934$ ($R^2=0.9998$); the power-of-two convention gives $p=0.935$ ($R^2=0.9997$). The tabulated energy ratio refers to the electric field only; total field energy is shown in Fig.~\ref{fig:energy}.}}
\begin{tabular}{lrrr}
\hline
$\CFL$ & Steps & Relative error &
$\mathcal{E}_{\text{ter}}/\mathcal{E}_{\text{fp}}$ \\
\hline
$0.1$      & $2000$  & $1.9437$ & $4.3098$ \\
$0.05$     & $4000$  & $1.0135$ & $1.6581$ \\
$0.025$    & $8000$  & $0.5162$ & $1.1011$ \\
$0.0125$   & $16000$ & $0.2804$ & $1.0018$ \\
$0.00625$  & $32000$ & $0.1451$ & $0.9834$ \\
\hline
\end{tabular}
\end{table}

The choice of Eq.~\eqref{eq:ntf}, rather than the canonical second-order form $N(z)=(1-z^{-1})^2$~\cite{Cal15}, is consequential. With the canonical form the encoder overloads persistently:~in the driven-cavity test the quantiser saturated on $49.0\%$ of updates and the error register was clamped on $40.5\%$, and the resulting ternary stream no longer provided a useful representation of the field. \RII{Equation~\eqref{eq:ntf} removes both. Over all $152\,100\,864$ encoder calls of the same test, that is over every component, cell and time step, the quantiser saturated $1759$ times ($1.16\times10^{-3}\%$) and the error register was clamped $348$ times ($2.29\times10^{-4}\%$), a reduction of more than four orders of magnitude in each.} While a first-order modulator $N(z)=1-z^{-1}$ can also help avoid overloading, it provides weaker in-band noise suppression:~in the same driven-cavity test, it produced a probe amplitude $1.8$ times smaller than that obtained with the second-order encoder.

\RIII{The encoder operates here in an unusual regime:~its input is not an external signal but the state of a dynamical system driven by its own output, so that the modulator and solver form a closed loop. Published stability results for sigma-delta modulators address first-order structures with constant or slowly varying inputs, as well as particular second-order structures under restrictive assumptions on the input~\cite{Cal15}. These results therefore do not directly apply to the configuration considered here and no analytical stability criterion is derived. Instead, the behaviour is characterised empirically, and how hard the encoder is driven depends on the excitation. Neither saturation nor clamping occurred at any $\CFL$ in the undriven convergence runs of Table~\ref{tab:convergence} presented below. (The rates for the driven test are those reported above.) The encoder parameters were chosen for the configuration studied here; retuning may be required for larger grids or heterogeneous media.}

The full-scale amplitude $A$ must exceed the largest field encountered anywhere on the Yee grid, since the encoder cannot represent more. For the configuration studied here a floating-point run gives $\max|\Ez| = 16.208$ and $\max|\mathbf{H}| = 7.619$, and I take $A=20$. Choosing $A$ too large is not free:~the quantisation error scales with $A$, so the tightest admissible value should be used. \RM{(On the other hand, instrumenting the accumulator shows that the scheme has its own dynamics. The requirement is therefore that $A$ exceed the maximum of the ternary field, which a floating-point pilot run underestimates by a factor that grows with $\CFL$. Although $A=20$ remains adequate, for problems whose peak field is not known in advance the choice of $A$ is an open question, discussed further below.)}

\begin{figure*}[t]
\includegraphics[width=\textwidth]{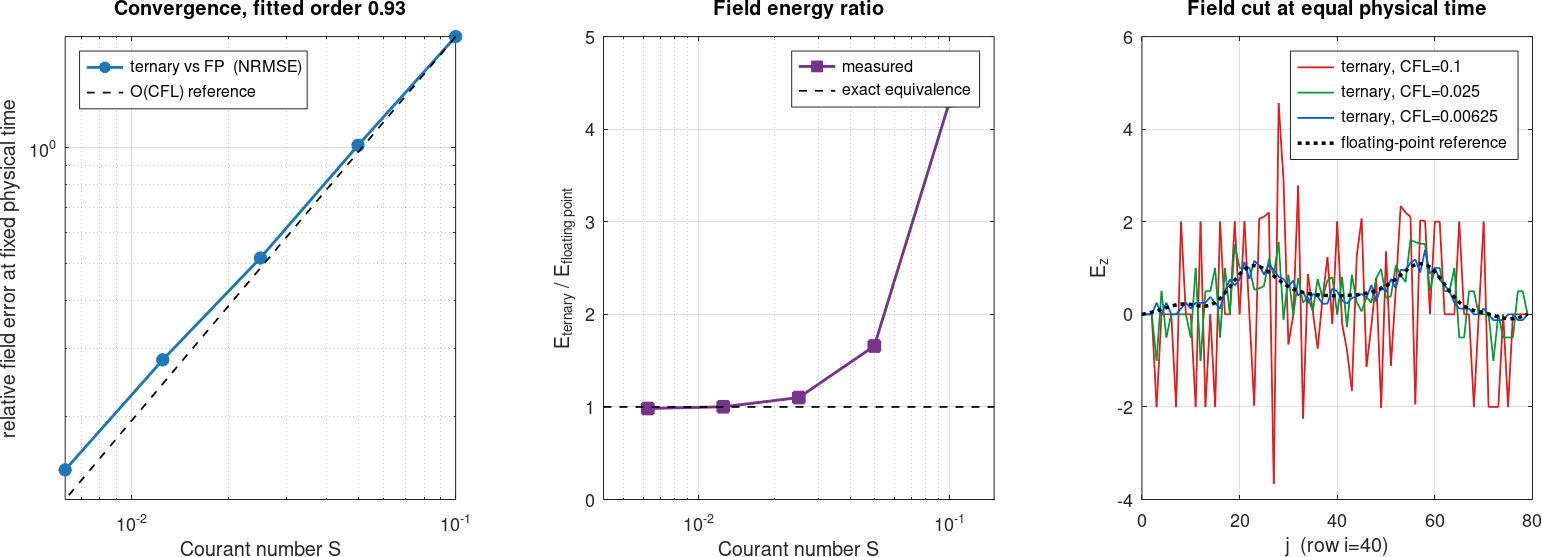}
\caption{\label{fig:convergence}Left:~relative error between the ternary \RII{state field $a^{E}A/M$} and the floating-point $\Ez$ field at fixed physical time, against the Courant number $\CFL$, with an $O(\CFL)$ reference line. Centre:~ratio of ternary to reference electric field energy, \RII{formed from the same two fields,} which falls from $4.31$ to within $10\%$ of unity by $\CFL=0.025$ and is flat thereafter. Right:~\RII{a cut through row $i=40$ at equal physical time for three values of $\CFL$, comparing the ternary state field $a^{E}A/M$ against the
floating-point $\Ez$ reference.} At $\CFL=0.1$ it obscures the field entirely; at $\CFL=0.00625$ it is reduced to a small oscillation about the reference, of amplitude consistent with the $0.145$ relative error of the left panel. \RII{The convergence order $p$ is obtained as the slope of an unweighted ordinary least-squares fit of $\log E_{\mathrm{rel}}$ against $\log\CFL$ over all five resolutions.}}
\end{figure*}

\subsection{\RII{Asymptotic agreement}}
The preceding sections establish what the scheme computes, but not yet in what sense it reproduces Eqs.~\eqref{eq:yee-hx}--\eqref{eq:yee-ez}. The key difficulty is that a ternary field cannot represent amplitude directly. The two schemes therefore cannot agree pointwise. The appropriate comparison must instead be made through the quantity encoded by the ternary symbols. The standard linearised analysis of a noise-shaped modulator provides precisely this framework, separating the encoded signal from the error introduced by its quantisation.

The encoder may be described, in the usual linearised noise model and away from quantiser overload, as
\begin{equation}
  s = \frac{x}{M} + q, \qquad q = N(z)\varepsilon\,,
  \label{eq:linear}
\end{equation}
where $\varepsilon$ is the raw rounding error, bounded by half a quantisation interval in the absence of clipping. The error-feedback structure shapes this error through $N(z)$, suppressing its low-frequency components. \RII{Equation~\eqref{eq:linear} applies only when the encoder is not overloaded. The full algorithm contains three nonlinear operations: rounding, ternary saturation and anti-windup clamping. The analysis neglects these effects by assumption, based on the measured overload rate reported above rather than on a formal bound. In particular, a bounded rounding error alone does not establish global stability of the nonlinear encoder, nor does it guarantee that its output spectrum is exactly described by $N(z)$.}

\RM{Substituting Eq.~\eqref{eq:linear} into the state updates and
multiplying through by $A/M$, with $M_C/M=\CFL$, separates each update into
two contributions. For the electric field,}
\begin{align}
  \Ez|_{i,j} &\leftarrow \Ez|_{i,j}
    + \CFL\Big[\left(\Hy|_{i,j}-\Hy|_{i-1,j}\right)\nonumber\\
  &\hspace{13mm}-\left(\Hx|_{i,j}-\Hx|_{i,j-1}\right)\Big]\nonumber\\
  &\quad\, + \CFL A\Big[\left(q^{H_y}|_{i,j}-q^{H_y}|_{i-1,j}\right)\nonumber\\
  &\hspace{13mm}-\left(q^{H_x}|_{i,j}-q^{H_x}|_{i,j-1}\right)\Big]\,,
  \label{eq:split}
\end{align}
\RM{and analogously for the two magnetic field components. The first part of Eq.~\eqref{eq:split} is Eq.~\eqref{eq:yee-ez} exactly. The second is a source absent from the reference scheme:~the ternary scheme is the Yee scheme perturbed by a bounded, spectrally shaped forcing}
\begin{equation}
  \mathbf{F}_{\text{err}} \sim \CFL A\,\nabla\times\left(N(z)\varepsilon\right)\,.
  \label{eq:forcing}
\end{equation}
\RII{The perturbed-update representation of Eq.~\eqref{eq:forcing}, together with the numerical results reported below, supports asymptotic agreement between the two schemes as $\CFL\rightarrow0$. However, Eq.~\eqref{eq:forcing} alone does not show that the accumulated error vanishes over the $O(\CFL^{-1})$ steps required to reach a fixed physical time:~a per-step perturbation of size $O(\CFL)$ could, in the absence of additional cancellation or stability estimates, accumulate to $O(1)$ over that interval. Noise shaping provides a plausible cancellation mechanism, while the convergence study in Table~\ref{tab:convergence} demonstrates the resulting behaviour numerically. Neither, however, constitutes a deterministic convergence theorem for the nonlinear, clamped error-feedback recursion, and no such theorem is claimed here. Rigorous stability and convergence analyses of sigma-delta modulators are available for important special cases, including first-order and constant-input settings, but these results do not directly establish convergence for the coupled dynamical system considered here~\cite{Cal15}.} Equation~\eqref{eq:forcing} also makes the associated cost explicit:~the perturbation scales with $\CFL A$ and can be reduced by refining the time step or reducing the full-scale amplitude.

\section{Courant Number as an Oversampling Ratio}
A three-level field can carry a physical wave through scale separation controlled by the Courant number. The 2D Yee dispersion relation is~\cite{Taflove}
\begin{equation}
  \sin^2\!\left(\frac{\omega\Delta t}{2}\right)
  = \CFL^2\left[\sin^2\!\left(\frac{k_x\Delta x}{2}\right)
              + \sin^2\!\left(\frac{k_y\Delta y}{2}\right)\right]\,,
  \label{eq:dispersion}
\end{equation}
which, measuring frequency in cycles per step so that
$\omega\Delta t=2\pi f$, gives
\begin{equation}
  f = \frac{1}{\pi}\arcsin\!\left(\CFL\sqrt{
      \sin^2\!\left(\frac{k_x\Delta x}{2}\right)
    + \sin^2\!\left(\frac{k_y\Delta y}{2}\right)}\right).
  \label{eq:fgeneral}
\end{equation}
The principal branch gives the positive-frequency solution in the Nyquist interval $0\leq f\leq1/2$. The arcsine argument must not exceed unity, requiring $\CFL\leq1/\sqrt{2}$, the 2D Courant stability condition~\cite{Taflove}.

The admissible wavenumbers follow from the PEC boundary conditions. With $\Ez$ forced to zero at $i=0$ and $i=N_x-1$, the cavity length is $(N_x-1)\Delta x$ and
$k_x=m\pi/[(N_x-1)\Delta x]$, with an analogous expression for $k_y$. Equation~\eqref{eq:dispersion} then yields the eigenfrequencies
\begin{equation}
f_{m,n}=\pi^{-1}\arcsin\Big(\CFL\sqrt{
\sin^2\tfrac{\pi m}{2(N_x-1)}+\sin^2\tfrac{\pi n}{2(N_y-1)}}\Big).
\label{eq:modes}
\end{equation}
Hence, every eigenmode supported by the grid lies below
\begin{equation}
  f_{\max} = \frac{1}{\pi}\arcsin\!\left(\CFL\sqrt{2}\right)
  \;\simeq\; \frac{\sqrt{2}}{\pi}\,\CFL\,.
  \label{eq:fmax}
\end{equation}
Even the highest-frequency cavity mode lies extremely close to this limit:~for $(m,n)=(N_x-2,N_y-2)$, I obtain $f=0.045158$, compared with $f_{\max}=0.045167$ at $\CFL=0.1$. Thus, all $(N_x-2)(N_y-2)=6084$ cavity modes lie within the narrow frequency range $[8.95\times10^{-4},\,0.04517]$, far below the Nyquist frequency of $0.5$. Frequencies above $f_{\max}$ cannot correspond to physical modes supported by the discrete wave operator and therefore provide spectral space into which quantisation noise can be shaped.

The encoder, meanwhile, places its error preferentially near the Nyquist frequency:~the noise transfer function of Eq.~\eqref{eq:ntf} vanishes at $f=0$ and is largest at
$f=1/2$. The two facts together define an oversampling ratio
\begin{equation}
  \mathrm{OSR} = \frac{1/2}{f_{\max}} \simeq \frac{\pi}{2\sqrt{2}\,\CFL}
  \approx \frac{1.11}{\CFL}\,,
\end{equation}
which is approximately $11$ at $\CFL=0.1$ and $178$ at $\CFL=0.00625$. Thus, reducing the Courant number does not only refine the time step but also widens the gap between the band the physics occupies and the band into which the encoder has been instructed to push its error.

\section{Numerical results}
\RII{A numerical convergence test of asymptotic agreement} between the ternary and \RII{floating-point reference} schemes is a convergence study. Both solvers are initialised with the same 2D Gaussian pulse $E_z=10\exp\left(-r^2/50\right)$ centred in the $80\times80$ cavity, where $r$ is the distance from the centre and the Gaussian standard deviation is $5$ cells. In the ternary case, the state accumulators are loaded with its fixed-point representation and encoded. \RII{The state accumulators therefore reproduce the initial field to within $A/2M$, while the emitted symbols encode it in the sense of Eq.~\eqref{eq:linear}.} The pulse enters as an initial condition and no source is applied thereafter, so the only forcing acting on the ternary field over the run is the quantisation term of Eq.~\eqref{eq:forcing}. Each run is propagated for the same physical time, the step count scaling as $\CFL^{-1}$.

\begin{figure*}[t]
\includegraphics[width=\textwidth]{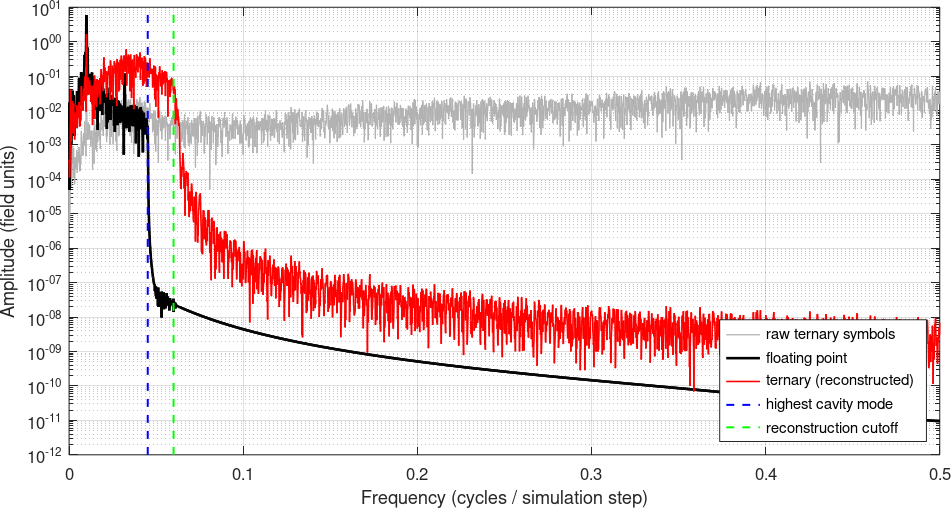}
\caption{\label{fig:spectrum}Full-band amplitude spectra at the probe cell, $\CFL=0.1$. Grey, the raw ternary symbol stream; black, the floating-point reference; red, the reconstructed ternary field. Dashed lines mark the highest frequency the grid can support, $f_{\max}=0.04517$ from Eq.~\eqref{eq:fmax} (blue) and the reconstruction cutoff $f_c=0.06$ (green). The raw stream is nearly white noise at $\sim2\times10^{-2}$ field units across four fifths of the spectrum, where the discrete wave operator has no support at all; only $2.9\%$ of its power lies below $f_{\max}$. \RII{Low-pass reconstruction removes the remainder:~the mean ratio of raw to reconstructed amplitude, evaluated over the band $f>0.1$ which contains $96.5\%$ of the raw stream power, is $1.4\times10^{4}$.} What survives is the in-band fraction, visible as the red plateau lying roughly one decade above the reference between $f_0$ and $f_{\max}$.}
\end{figure*}

As shown in Table~\ref{tab:convergence} and Fig.~\ref{fig:convergence}, halving $\CFL$ approximately halves the error, with a fitted convergence order of $0.93$ over a factor of $16$ in $\CFL$, while the energy ratio approaches unity from above. This provides the substantive \RII{asymptotic-agreement} result: the ternary scheme is consistent with the Yee scheme and converges with approximately first-order accuracy in the Courant number.

\subsection{Spectral Analysis}
I consider an $80\times80$ PEC-wall cavity driven sinusoidally at $f_0=0.01$ for $8192$ time steps at $\CFL=0.1$, with the response measured at a single $(32,32)$ probe cell. Figure~\ref{fig:spectrum} shows that the ternary scheme reproduces the excitation frequency of the \RII{floating-point reference} to within one Fourier-transform bin.

The amplitude at this frequency, however, is $28\%$ of the reference value because most of the power in the ternary signal is not at the physical frequencies supported by the cavity. Instead, $96.5\%$ lies above $f=0.1$, beyond the physical bandwidth of the grid given by Eq.~\eqref{eq:fmax}. This power is quantisation noise generated by the ternary representation. The encoder deliberately moves this noise towards high frequencies, while the reconstruction filter removes it. Below the filter cutoff, the reconstructed ternary spectrum follows the reference closely. Above the cutoff, both spectra are strongly attenuated. The higher residual floor in the ternary spectrum therefore reflects quantisation residue rather than additional physical modes.

Although the driving line is more than an order of magnitude above the local noise floor, the residual quantisation noise is distributed broadly across hundreds of Fourier bins. This observation does not indicate a failure to reproduce a correct physical response, since the excitation remains clearly identifiable as a coherent spectral component. Rather, the broadband numerical noise \RII{produces an effective reduction in coherent modal accumulation} in the lossless cavity, continually exciting non-target modes and thereby limiting the accumulation of coherent energy in the driven mode.

In fact, substantial in-band noise is expected at $\CFL=0.1$, since the corresponding oversampling ratio is $\mathrm{OSR}\simeq11$ and the physical signal band is separated from the quantisation-noise floor only moderately. Decreasing $\CFL$ reduces the amplitude of each injected quantisation input, proportional to $\CFL A$, while simultaneously increasing the frequency range between the physical cutoff $f_{\max}$ and the Nyquist frequency. These effects reduce the in-band numerical noise and provide the mechanism for the convergence observed in Table~\ref{tab:convergence}.

\begin{table}[t]
\caption{\RIII{\label{tab:driven}Driven-mode amplitude against the Courant number $\CFL$, at fixed physical time and fixed physical drive frequency. Both the drive frequency in cycles per step and the soft-source increment are scaled with $\CFL$, so that the physical frequency and the injected power per unit physical time are invariant. The amplitude is obtained by coherent demodulation of the raw symbol stream over the second half of each run, with no reconstruction filter applied.}}
\begin{tabular}{lrrrr}
\hline
$\CFL$ & Steps & $A_{\text{ter}}$ & $A_{\text{ref}}$ & Error \\
\hline
$0.10000$ & $8192$   & $2.791$ & $8.450$ & $66.98\%$ \\
$0.05000$ & $16384$  & $3.668$ & $8.332$ & $55.98\%$ \\
$0.02500$ & $32768$  & $5.554$ & $8.302$ & $33.11\%$ \\
$0.01250$ & $65536$  & $6.926$ & $8.294$ & $16.50\%$ \\
$0.00625$ & $131072$ & $7.589$ & $8.293$ & $8.48\%$  \\
$0.00312$ & $262144$ & $7.935$ & $8.296$ & $4.36\%$  \\
\hline
\end{tabular}
\end{table}

\subsection{\RIII{Driven-mode accuracy against the Courant number}}
\RIII{The $28\%$ amplitude deficit reported above is a property of the oversampling ratio rather than of the ternary representation, and it converges away. To establish this, and to quantify how small $\CFL$ must be before the driven-mode amplitude error falls below a prescribed tolerance, the driven cavity was run at six Courant numbers spanning a factor of $32$, in each case at fixed physical time and fixed physical drive frequency.}

\RIII{Two normalisations are required for such a sweep to be meaningful. First, the drive frequency measured in cycles per step must scale with $\CFL$ so that the physical frequency is invariant. Second, the soft source must inject per unit physical time rather than per time step (without this, halving $\CFL$ while doubling the step count doubles the total injected energy). The reference amplitude is reported in Table~\ref{tab:driven} precisely as the control on this point, and its variation of $1.89\%$ across a factor of $32$ in $\CFL$ confirms that the comparison is at fixed physical excitation.}

\begin{figure*}[t]
\includegraphics[width=\textwidth]{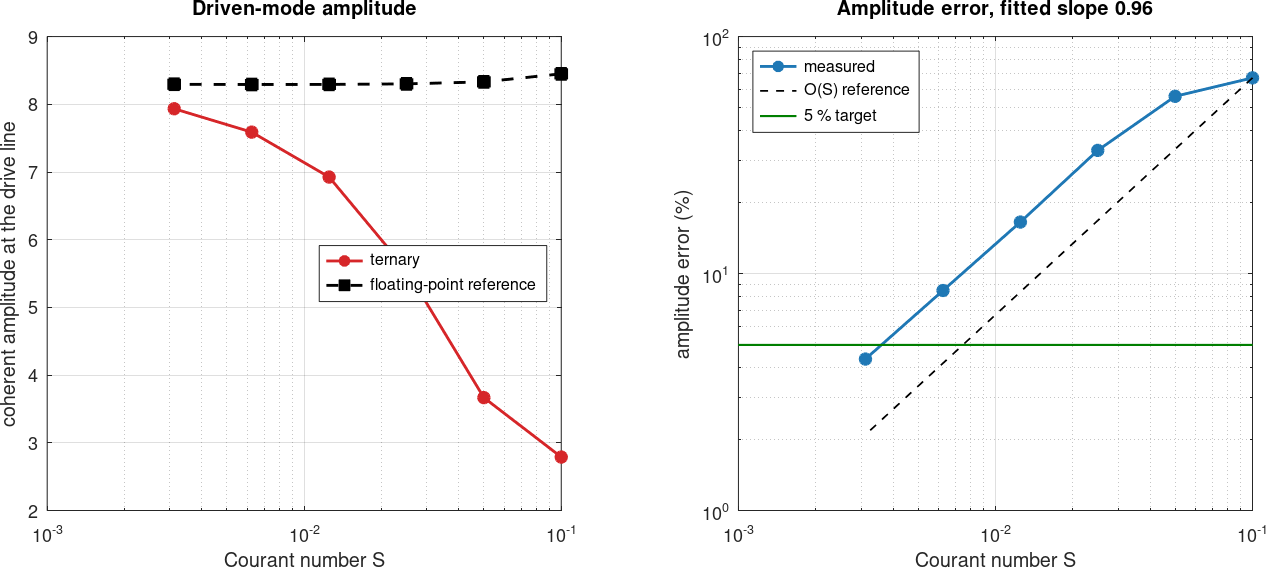}
\caption{\RIII{\label{fig:driven}Left:~coherent amplitude at the drive line against the Courant number $\CFL$, formed at the probe cell from the raw ternary symbols $sA$ and from the floating-point $\Ez$. The near-constancy of the reference curve is the control confirming fixed physical excitation. Right:~the corresponding amplitude error on log--log axes, with an $O(\CFL)$ reference
line anchored to the coarsest point and the $5\%$ target marked. The error falls essentially linearly in $\CFL$, crossing $5\%$ at $\CFL=0.0036$.}}
\end{figure*}

\RIII{Figure~\ref{fig:driven} visualises the results presented in Table~\ref{tab:driven}. The amplitude deficit falls monotonically, with a fitted slope of $0.96$ in log--log over the three finest points, i.e.~essentially first order in $\CFL$ and consistent with the field-error convergence of Table~\ref{tab:convergence}. The $5\%$ threshold is crossed between $\CFL=0.00625$ and $\CFL=0.00312$; log--log extrapolation gives $\CFL=0.0036$. The corresponding computational penalty is $28$ times the step count at $\CFL=0.1$ and $196$ times the step count of a conventional solver running at the stability limit $\CFL=1/\sqrt2$.}

\RIII{This numerical analysis also reveals that the amplitude is obtained by projecting the raw symbol stream onto a complex exponential at the drive frequency and that no low-pass filter of any kind is involved. It is therefore a modal amplitude measurement at a physical resonance that is independent of every element of the comparison methodology described below, and it provides the physically feasible, filter-free benchmark.}

\subsection{\RIII{Long-time energy behaviour}}
\begin{figure*}[t]
\includegraphics[width=\textwidth]{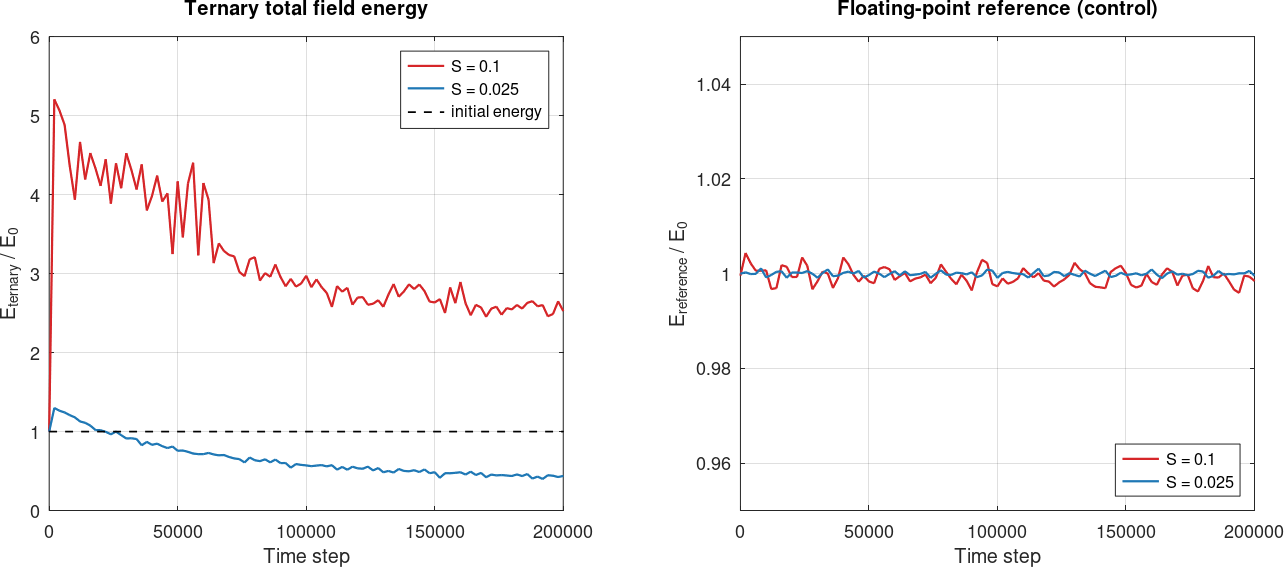}
\caption{\RIII{\label{fig:energy}Total field energy, including all three components, over $2\times10^{5}$ steps of an undriven Gaussian pulse, normalised to the initial value. Left:~ternary scheme at two Courant numbers. The energy rises over an initial transient, saturates, and then drifts slowly downward; it is bounded, not runaway. Right:~the floating-point reference over the same runs, which conserves energy to within $0.43\%$ at $\CFL=0.1$ and $0.11\%$ at $\CFL=0.025$, and which is the control that makes the left panel interpretable.}}
\end{figure*}

\RIII{The energy ratios of Table~\ref{tab:convergence} raise the question of whether the scheme is energy-stable over long integrations or whether the drift observed at moderate $\CFL$ eventually grows without bound. To address this question, an undriven Gaussian pulse was released at the centre of the cavity and both schemes were integrated for $2\times10^{5}$ steps, one hundred times the duration of the convergence experiments of Table~\ref{tab:convergence} and corresponding to $2\times10^{4}$ units of physical time at $\CFL=0.1$. Figure~\ref{fig:energy} shows the total field energy, normalised to its initial value.}

\RIII{The behaviour is not unbounded growth. At $\CFL=0.1$ the ternary energy rises to $5.20$ times its initial value over a transient of order $2000$ steps, saturates and thereafter drifts slowly downward at $-0.29$ of the initial energy per $10^{5}$ steps, ending at $2.53$ with a late-time mean of $2.67\pm0.13$. At $\CFL=0.025$ the peak is $1.30$, the late-time mean is $0.488\pm0.048$ and the drift is $-0.15$ per $10^{5}$ steps; the late-time energy lies below the initial value. The floating-point reference conserves energy to within $0.43\%$ and $0.11\%$ respectively over the same intervals, which is the control that makes the ternary curves interpretable.}

\RIII{There is a structural reason to expect boundedness. The encoder error registers are hard-clamped to $\pm2M$ at every invocation, so the forcing injected per step is bounded by a constant independent of the field amplitude, whereas the coupling that returns energy to the quantiser grows with the field. The scheme cannot sustain unbounded growth for the same reason that a sigma-delta modulator with bounded integrator state cannot:~the error injection does not scale with the signal. I emphasise that this is a bounded-in-practice demonstration over a long but finite integration together with a structural argument, and not a formal discrete energy estimate, which is not derived here.}

\subsection{\RM{Computational cost}}
\RM{The motivation for this work stems in part from ternary large language models designed for inference on mobile and resource-constrained hardware, where reduced computational and memory requirements can lower energy consumption and inference time~\cite{Ma24, Qia26}. However, ternary quantisation restricts model parameters to a small discrete set, slowing convergence relative to floating-point models~\cite{Den18}. Nevertheless, ternarisation enables increasingly capable large language models to run on edge devices, reducing reliance on expensive datacentre GPUs and cloud infrastructure~\cite{dade2026}.

The main computational cost of the ternary FDTD scheme is the smaller time step required for accurate results. I first compared the two update loops within the same binary, using identical hardware, compiler, optimisation flags, grid and number of steps. On a general-purpose processor, one ternary update takes $13.6$ times longer than the corresponding single-precision floating-point update. This difference reflects the instruction count:~one $\Ez$ cell update requires $53$ instructions for the ternary scheme, compared with $9$ for the floating-point reference. The smaller time step introduces an additional cost. According to the driven-mode results in Table~\ref{tab:driven}, achieving a $5\%$ amplitude error requires $28$ times more steps than at $\CFL=0.1$, and $196$ times more steps than a conventional solver operated at its stability limit.}

\RM{These CPU timings provide a useful reproducible baseline, but they do not represent the intended hardware advantage of the ternary formulation. General-purpose processors are designed to execute floating-point arithmetic efficiently, whereas the ternary update uses mainly additions, comparisons and fixed shifts. The CPU therefore provides a conservative test of the arithmetic itself rather than of the hardware efficiency that the formulation is intended to enable.}

\RM{The potential advantage becomes more relevant in hardware architectures where arithmetic resources and memory bandwidth constrain the number of cells that can be updated in parallel. The ternary representation requires only two bits per field component, compared with $32$ bits for single-precision floating-point data, and eliminates multiplication and floating-point normalisation and exponent handling. On an FPGA, this can reduce logic and memory requirements and allow more stencil updates to be instantiated in parallel. On processors without hardware floating-point support, the balance can likewise differ because floating-point operations may require software emulation, whereas the ternary update can use native integer operations. The resulting trade-off between arithmetic cost, memory use, parallelism and the increased step count is therefore architecture-dependent. A detailed hardware evaluation is beyond the scope of the present work and is addressed in the companion implementation study~\cite{Maksymov2026TernaryVHDL}.}

\section{Comparison Methodology}
A ternary field is not a coarsely rounded field and treating it as such introduces artefacts that can obscure the underlying behaviour. The following measurement choices are therefore necessary.

\subsection{Demodulation before comparison}
The sequence of ternary symbols carries its information in local symbol density, not in instantaneous values, and by construction places its error near the Nyquist frequency. A direct sample-by-sample comparison against the reference is therefore conducted almost entirely within the error band and returns a correlation near zero regardless of whether the underlying field is correct. Consequently, in the time domain I use a linear-phase finite-impulse-response low-pass with cutoff above $f_{\max}$ and well below the shaped noise, applied to the reference as well as to the symbol stream so that both are compared through the same transfer function. (Recall that the reconstructed ternary field is $A$ times the filtered stream.) In the frequency domain, similarity measures computed over the full spectrum are dominated by the shaped error, so I restrict spectral analysis to $f\leq f_{\max}$ from Eq.~\eqref{eq:fmax}. Spatial comparison is made without filtering, by the polarity scoring.

\RII{The reconstruction operator is a $401$-tap finite impulse response low-pass filter formed by truncating an ideal sinc response to $401$ samples and applying a symmetric Hann window. The ideal sinc prototype has $f_c=0.06$ cycles per step, defined at its $-6$~dB point. The resulting coefficients are normalised to unit gain at zero frequency by division by their sum. Filtering is single-pass and linear-phase, with a constant group delay of $200$ samples removed by symmetric alignment of the convolution. No forward--backward (filtfilt-type) filtering is used, as this would square the magnitude response and therefore change the stopband attenuation reported in Fig.~\ref{fig:spectrum}. The first and last $200$ output samples are discarded because of the finite-record boundary transient, together with a further $600$ samples at the beginning to remove the startup transient, leaving $7192$ of the original $8192$ samples for spectral analysis.}

\subsection{Polarity scoring against a shuffled-symbol null}
For simulated ternary snapshots, it is natural to ask how often the ternary symbol agrees in sign with the reference field. Raw polarity agreement has no natural zero point and must be referred to a control. The control I adopt is a reshuffle:~the emitted symbols are kept exactly as the computation produced them, the counts of $+1$, $-1$ and $0$ held at their measured values, and are redistributed uniformly at random over the grid, so that every property of the symbol population survives and only its spatial arrangement is destroyed. The agreement obtained against this control is the agreement attributable to the symbol statistics alone and I denote it $p_{\text{chance}}$.

Its value follows from the marginal frequencies. Under the reshuffle a symbol is independent of the cell it occupies, so if a fraction $f_+$ of the usable reference cells is positive and a fraction $t_+$ of the emitted symbols is
$+1$, the two signs agree with probability
\begin{equation}
  p_{\text{chance}} = f_+ t_+ + (1-f_+)(1-t_+)\,,
  \label{eq:chance}
\end{equation}
which reduces to one half whenever either marginal population is balanced, i.e.~$f_+=1/2$ or $t_+=1/2$.

For the encoder used here, $t_+=0.499$, so Eq.~\eqref{eq:chance} lies within $0.2\%$ of one half for any value of $f_+$. The floor is nevertheless determined empirically, for two reasons. First, Eq.~\eqref{eq:chance} assumes that symbol occupancy is independent of position, whereas emission on the numerical grid is suppressed where the reference amplitude is small, prohibited inside the conductor and modulated by the standing-wave pattern elsewhere. Second, the near balance of $t_+$ is a property of this operating point rather than of the encoder construction (an encoder driven closer to saturation is not guaranteed to retain this balance).

The control preserves the symbol population and destroys only its spatial arrangement. The counts of $+1$, $-1$ and $0$ are held at their measured values, and the symbols are redistributed uniformly at random over the grid, then scored against the reference by the same procedure applied to the field itself. The resulting agreement is therefore attributable to the symbol statistics alone.

\RII{A cell contributes to the polarity score only if its ternary symbol is non-zero and the magnitude of the reference field is at least $10\%$ of the maximum absolute reference-field magnitude in the frame. A zero ternary symbol makes no polarity claim, while a reference field close to a zero crossing does not provide a reliable polarity. The same exclusions are applied to the numerator and denominator, and identically to the shuffled control, so that both the measured agreement and its shuffled baseline are evaluated over the same cell population. The shuffled baseline is estimated from $200$ independent reshuffles.}

\begin{figure*}[t]
\includegraphics[width=0.9\textwidth]{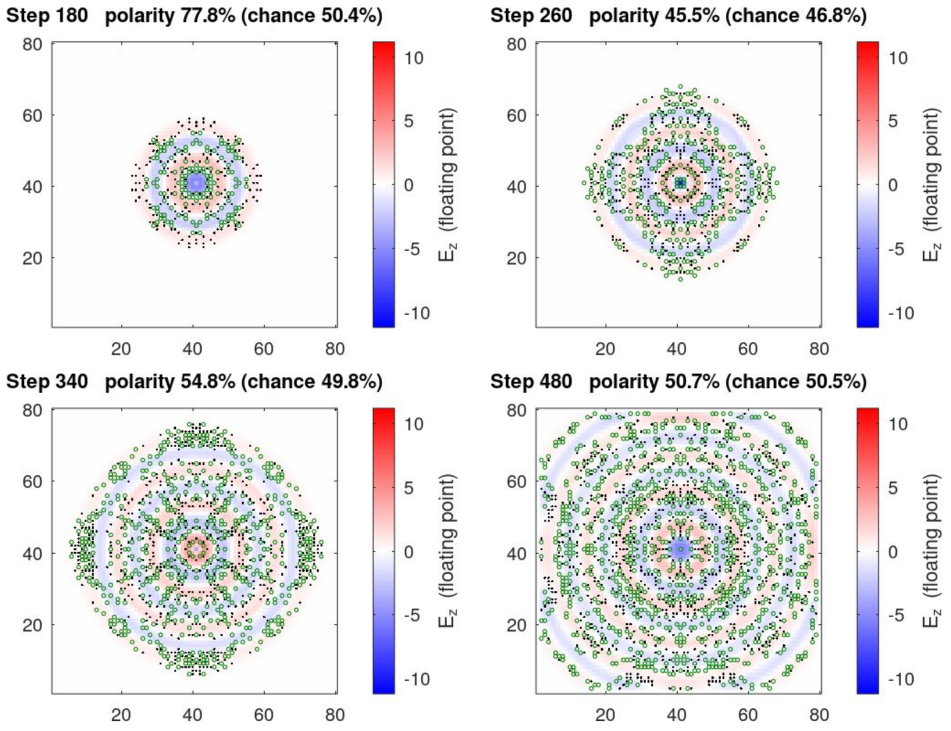}
\caption{\label{fig:polarity}Ternary symbols overlaid on the floating-point $E_z$ field at four steps of the simulation, $\CFL=0.1$. Black dots mark cells emitting $+1$, green circles cells emitting $-1$. The colour map is the reference field on a shared scale. Panel titles give the polarity agreement $p$ and the shuffled-symbol floor $p_{\text{chance}}$\RII{, the latter estimated from $200$ reshuffles, which fixes it to better than $0.6\%$ in every panel.}}
\end{figure*}

Figure~\ref{fig:polarity} overlays the emitted symbols on the reference floating-point field at four steps of the simulation. The spatial organisation is evident by inspection:~symbols are confined to the wave region, absent from the quiescent grid ahead of the wavefront and arranged in concentric annuli that coincide with the fringes of the reference field. The expanding envelope tracks the reference, which is a direct confirmation that the ternary scheme propagates at the correct numerical velocity. Symbol density also follows the local amplitude, sparse in the outer rings and saturating near the source, which is the behaviour expected of a modulator whose output rate encodes magnitude.

The quantitative scores separate into two regimes. At step $180$ the wavefront is compact, the amplitude per illuminated cell is large and agreement reaches $77.8\%$ against a floor of $50.6\%$. As the field spreads over the full grid the energy per cell falls, the encoder spends an increasing fraction of its time near zero crossings and the scores decline towards the floor. \RII{At steps $340$ and $480$, the agreement is $54.9\%$ and $53.2\%$ against floors of $49.9\%$ in both cases.} Step $260$ falls marginally below chance, at $45.5\%$ against $50.9\%$. I attribute this to the memory held in the encoder error registers, which carries the polarity of the preceding half-cycle across a field reversal. The four frames sample different phases of the $100$-step drive period and step $260$ lies close to a global sign change. \RII{With $264$ cells entering the score at this step, the sampling error on polarity agreement $p$ is $3.1$ percentage points and the measured $p$ lies $1.3$ standard errors below the floor, which is consistent with chance rather than indicating a systematic reversal.}

This decline is the expected behaviour of the representation rather than a deficiency of the scheme. A sigma-delta stream carries its information in the local average of the emitted symbols, with symbol density providing a direct measure of that average, rather than in the instantaneous sign of any individual cell. The encoder can therefore emit symbols even when the instantaneous amplitude is below one quantisation quantum, depending on the state of its error accumulator; a cell adjacent to a zero crossing may fire in either direction according to the accumulated error.

\RII{Second-order noise shaping in fact requires opposite-sign correction pulses, since that is the mechanism by which it displaces error towards high frequency. A substantial fraction of the emitted symbols at any instant therefore carry the locally wrong instantaneous sign while being correct behaviour, and the attainable per-cell agreement is bounded well below unity.} Per-cell polarity is thus a deliberately stringent observable, since it discards the spatial averaging on which the encoding depends. That a compact, well-illuminated wavefront nonetheless reaches $77.8\%$ against a floor of about $50\%$ shows that substantial spatial information survives in the raw symbols before any reconstruction is applied\RII{ and may be close to what this measure can give}. \RIII{Its role here is to demonstrate spatial organisation. The quantitative accuracy of the scheme is established by the convergence study of Table~\ref{tab:convergence} and by the filter-free driven-amplitude measurement of Table~\ref{tab:driven}.}

\section{Discussion}
The proposed update requires no run-time multiplication beyond compile-time constants \RII{when $M$ and $\CFL$ are chosen as powers of two}, no floating-point arithmetic and represents each \RII{emitted} field \RII{symbol} with the theoretical $\log_2 3\simeq1.58$ bits limit, requiring only two physical bits in a binary implementation~\cite{Ma24, Qia26}.

\RM{The two-bit figure describes the emitted symbol and the memory of the scheme is reorganised rather than simply reduced. Each cell additionally owns, for every field component, an integer state accumulator and two encoder-error registers. Measuring the required word lengths rather than estimating them gives $21$ bits signed for the accumulator, a factor of four of headroom above the observed maximum, and $19$ bits signed for each error register, bounded by the anti-windup clamp at $2M$. Complete per-component, per-cell storage is therefore about $61$ bits, against $32$ bits for a single-precision reference. What the ternary representation reduces is not storage but communication:~the accumulator and both registers are strictly cell-local, read and written only by the cell that owns them, so the sole quantity crossing a cell boundary is the two-bit symbol, sixteen times less than the thirty-two bits a floating-point stencil must exchange. Where an update is limited by data movement between processing elements rather than by local register-file capacity, this is the figure that governs performance.}

\RII{All state in the reported simulations is held in 32-bit signed integers. The largest values observed over all $152\,100\,864$ encoder invocations of the driven benchmark are $|a|=179\,259$, $|v|=380\,634$ and $|e_1|=280\,634$ before clamping, so that \texttt{int32\_t} provides a margin of more than 5000 over the largest quantity that arises, and no overflow is possible for this benchmark. The margin is problem-dependent for the reasons given in connection with the full-scale amplitude.}

\RII{The scheme is physically consistent and numerically convergent, and in that sense the ternary and floating-point solvers agree asymptotically.} The ternary-to-Yee error converges approximately first order in $\CFL$, whereas the underlying Yee discretisation is second order in both $\Delta x$ and $\Delta t$. Since the ternary error is controlled by $\CFL$ rather than by grid refinement, and because the step count scales as $\CFL^{-1}$ at fixed physical time, the computational cost of a given accuracy scales as
$\varepsilon^{-1}$ in steps, on top of the cost of the grid itself. At
$\CFL=0.1$, the quantisation error exceeds the field. \RIII{The scheme buys its arithmetic simplicity with time steps. Useful accuracy in the configuration studied here demands $\CFL\lesssim0.01$ by the field-error criterion and $\CFL\simeq0.0036$ by the driven-amplitude criterion, i.e.~$71$ to $196$ times more steps than a conventional solver running at its stability limit.}

\subsection{\RM{Selection of the full-scale amplitude}}
\RM{The full-scale amplitude $A$ must be large enough to accommodate the largest field magnitude reached anywhere on the grid, while remaining as small as possible because the quantisation error scales with $A$. A value estimated from a floating-point pilot run does not necessarily provide sufficient headroom: in the benchmark considered here, the ternary state reached a peak $2.21$ times larger than the maximum of the floating-point reference. Thus, for problems in which the peak field cannot be estimated reliably in advance, including nonlinear or strongly scattering configurations, the choice of $A$ presents a practical design constraint. Possible approaches include deriving an a priori bound on the ternary state, using adaptive rescaling with a specified overflow policy, or introducing additional headroom at the cost of increased quantisation error. This limitation is relevant when the method is used as a self-contained solver. In the scattering configuration discussed in the following, however, the same value of $A$ remained adequate and the saturation rate was unchanged, indicating that scattering alone does not necessarily require rescaling.}

\subsection{\RIV{Domain of applicability of the spectral-gap argument}}
\RIV{The construction relies on a separation between the frequencies occupied by the physical solution and the Nyquist frequency of the time sampling. It is therefore useful to state the conditions under which this spectral gap exists. The analysis above was developed for a homogeneous, translationally invariant Yee grid, where the discrete modes can be represented by plane waves, but the underlying argument is more general.}

\RIV{For a linear leapfrog discretisation of a wave system whose spatial operator admits a real, bounded spectrum, the relevant condition is that the highest physical temporal frequency remain below the Nyquist frequency. Writing as $\rho$ the largest eigenvalue of the discrete spatial operator whose symbol is $\sin^{2}(k_x\Delta x/2)+\sin^{2}(k_y\Delta y/2)$, the corresponding leapfrog dispersion relation gives
\begin{equation}
  f_{\max}=\frac{1}{\pi}\arcsin\!\left(\CFL\sqrt{\rho}\right)\,.
  \label{eq:fmaxrho}
\end{equation}
For the homogeneous 2D Yee grid, $\rho=2$, recovering Eq.~\eqref{eq:fmax}. Thus, plane-wave analysis provides a convenient way of obtaining the bound for the homogeneous grid, but more generally the existence of a spectral gap can be assessed from the spectrum of the discrete spatial operator.}

\RIV{For the hard PEC boundaries considered here, the admissible discrete modes are restricted by the boundary conditions and do not introduce frequencies above the corresponding homogeneous-grid bound. The eigenvalues of the finite-domain problem in Eq.~\eqref{eq:modes} satisfy this bound to within $10^{-5}$, as noted above. More general absorbing or impedance boundaries require greater care:~boundary operators can modify the spectrum and may produce non-normal or complex eigenvalues~\cite{Sau13}, so the homogeneous spectral-gap
argument cannot be assumed to remain unchanged without analysing the
particular boundary discretisation.}

\RIV{Heterogeneous media likewise require a separate spectral assessment. The relevant upper bound is determined by the largest eigenvalue of the variable-coefficient spatial operator, for which the maximum local wave speed provides a conservative CFL constraint but does not, in general, determine the spectrum by a simple scaling factor. A smaller time step can therefore increase the available separation from the Nyquist frequency, but material interfaces can substantially alter the modal structure and field distribution. In addition, the amplitude scale $A$ is determined by the largest field magnitude anywhere in the domain and is therefore a separate dynamic-range constraint. High material contrast may make this amplitude requirement more restrictive even when the spectral gap remains adequate. The spectral-gap and amplitude-scale requirements should therefore be regarded as distinct conditions.}

\RIV{For coupled first-order wave systems advanced with a Yee-type leapfrog scheme, the same argument can be extended when the coupled spatial operator has a bounded spectrum and the fields share a common time step. Virieux-type elastodynamic schemes~\cite{Vir84} provide an example of this class and are therefore a natural prospective extension. The time-dependent Schr{\"o}dinger equation does not fall directly into this class~\cite{Sullivan, Mak_FDTD}:~it is first order in time and involves a complex-valued field, so its spectral relation to the Nyquist frequency and the corresponding noise-shaping construction would need to be derived separately rather than inherited from the present analysis.}

\subsection{\RIII{Extension to three dimensions}}
\RIII{Considering an extension to a three-dimensional model, it is useful to note that the dispersion relation acquires a third sine term, so that $f_{\max}=\pi^{-1}\arcsin(\CFL\sqrt3)$ and the oversampling ratio becomes $\mathrm{OSR}\simeq0.91/\CFL$. The separation of scales on which the method depends is therefore slightly reduced but not qualitatively altered. Six field components each require their own accumulator and error-register pair, so the storage accounting given above applies per component as before, and the staggered couplings are more numerous but structurally identical.}

\begin{algorithm}[h]
\caption{\label{alg:step}One ternary time step, with PEC scatterer.}
\begin{algorithmic}[1]

\Function{Encode}{$x$; $e_{1},e_{2}$ \textbf{by reference}}
  \State $v \gets x + (3e_{1} - e_{2})/2$
  \State $s \gets \operatorname{round}(v/M)$ clamped to $[-1,+1]$
  \State $e_{2} \gets e_{1}$; \quad $e_{1} \gets v - sM$
  \State clamp $e_{1}$ to $[-2M, 2M]$ \Comment{anti-windup}
  \State \Return $s$
\EndFunction

\Statex
\Statex \textit{Magnetic field}
\For{$0 \leq i < N_x-1$, \; $0 \leq j < N_y-1$}
  \State $a^{H_x}_{i,j} \mathrel{+}= M_C\,(s^{E}_{i,j}-s^{E}_{i,j+1})$
  \State $a^{H_y}_{i,j} \mathrel{+}= M_C\,(s^{E}_{i+1,j}-s^{E}_{i,j})$
  \State $s^{H_x}_{i,j} \gets \Call{Encode}{a^{H_x}_{i,j};\,
         e^{H_x}_{1,i,j},\, e^{H_x}_{2,i,j}}$
  \State $s^{H_y}_{i,j} \gets \Call{Encode}{a^{H_y}_{i,j};\,
         e^{H_y}_{1,i,j},\, e^{H_y}_{2,i,j}}$
\EndFor

\Statex
\Statex \textit{Electric field}
\For{$0 < i < N_x-1$, \; $0 < j < N_y-1$}
  \State $a^{E}_{i,j} \mathrel{+}= M_C\big[(s^{H_y}_{i,j}-s^{H_y}_{i-1,j})
         - (s^{H_x}_{i,j}-s^{H_x}_{i,j-1})\big]$
  \State $s^{E}_{i,j} \gets \Call{Encode}{a^{E}_{i,j};\,
         e^{E}_{1,i,j},\, e^{E}_{2,i,j}}$
\EndFor

\Statex
\Statex \textit{Source at $(s_x,s_y)$, with rounding residual $\rho$}
\State $u \gets (M/A)\,g(n) + \rho$
\State $a^{E}_{s_x,s_y} \mathrel{+}= \lfloor u \rceil$; \quad
       $\rho \gets u - \lfloor u \rceil$
\State $s^{E}_{s_x,s_y} \gets \Call{Encode}{a^{E}_{s_x,s_y};\,
       e^{E}_{1,s_x,s_y},\, e^{E}_{2,s_x,s_y}}$

\Statex
\Statex \textit{Boundaries and conductor}
\State $s^{E}_{i,j} \gets 0$ on all four grid boundaries
\ForAll{$(i,j)$ in the PEC scatterer}
  \State $s^{E}_{i,j},\; a^{E}_{i,j},\;
         e^{E}_{1,i,j},\; e^{E}_{2,i,j} \gets 0$
  \Comment{registers must be cleared too}
\EndFor

\end{algorithmic}
\end{algorithm}

Algorithm~\ref{alg:step} outlines one complete time step of the corrected scheme. The encoder registers $e_1$ and $e_2$ are per component and per cell. The source term $g(n)$ is injected with its own rounding residual $\rho$, so that the mean injected amplitude is exact even though each individual increment is an integer. Omitting this residual introduces a systematic amplitude error at small $\CFL$, where the per-step increment becomes comparable to unity. Full-scale amplitude $A$ must exceed the largest field magnitude anywhere on the numerical grid at any time, since the encoder saturates otherwise. However, it should exceed it by as little as possible, since the quantisation error is proportional to $A$. Fixed-point resolution $M$ must be large enough that $M_C = M\CFL$ is an exact integer for every Courant number of interest and large enough that the state resolution $A/M$ is negligible against the field. It does not otherwise
affect accuracy, since the dominant error is the ternary encoding rather than the state representation. I use $M=10^5$, giving a state resolution of $2\times10^{-4}$ field units at $A=20$\RII{, and $M=2^{17}$ where a multiplication-free datapath is required, giving $1.5\times10^{-4}$ field units at the same $A$}.

\section{Conclusions}
A ternary FDTD scheme can reproduce the Yee algorithm, but not in the form in which it is most naturally written\RII{:~a quantiser that drains its own accumulator removes the temporal integration from the update and leaves an algebraic system supporting no propagating wave}. The presented algorithm separates the state, which integrates and is never drained, from a noise-shaped encoder carrying its own error registers, and introduces the Courant number explicitly as an exact fixed-point ratio. In such a construction, $S$ acquires a second role, as the parameter setting the oversampling ratio of the encoder. Every eigenmode the grid supports lies below $\arcsin(S\sqrt{2})/\pi$, while the shaped quantisation error is concentrated near the Nyquist frequency, so that $\mathrm{OSR}\simeq1.11/S$. \RII{Agreement with the Yee scheme is then asymptotic in $S$, with the relative field error at fixed physical time falling as $S^{0.93}$ and the field-energy ratio approaching unity. The result is a consistency statement supported by a convergence study rather than a theorem.}

\RIII{The scheme is characterised quantitatively rather than only in principle. On a general-purpose processor, the ternary update runs $13.6$ times slower per step than the single-precision reference and requires $71$--$196$ times more steps at matched accuracy. A processor that already provides hardware multiplication and floating-point arithmetic is therefore not the architecture for which the ternary formulation is intended, and no wall-clock advantage is claimed on such processors. Over more than $2\times10^{5}$ steps, the field energy saturates and then drifts slowly downward, demonstrating bounded long-time behaviour of the coupled encoder--solver system in the tested regime. The update itself requires only additions, comparisons and shifts, while each field component requires two bits of inter-cell communication and no multiplier or floating-point unit. This arithmetic and communication structure is directly suited to resource-constrained hardware, particularly FPGAs and embedded processors, where eliminating multipliers, floating-point units and wide data paths can increase the number of stencil updates that can be instantiated and operated in parallel~\cite{Mansoori, Mak_FDTD}. The present work therefore establishes the numerical accuracy, convergence and long-time behaviour needed to evaluate the corresponding hardware implementation; the achievable performance and energy efficiency are hardware-dependent quantities to be established by implementation.}

Algorithm~\ref{alg:step} extends to scattering problems without modification~\cite{Maksymov2026TernaryFDTD}. A PEC scatterer is imposed on the ternary field exactly as on the reference, by forcing $E_z$ to zero on the occupied cells. The only additional requirement is that the state accumulator and both encoder error registers be cleared alongside the emitted symbol, since leaving them to integrate inside the conductor would wind them against their anti-windup limits and store a latent charge with no counterpart in the reference. For instance, with a $9\times9$ square obstacle of side $0.9\lambda$ placed in the upper-right quadrant of the numerical grid, the ternary solver reproduces the expected illuminated and shadowed regions. The encoder remains unaffected.

\RM{For this demonstration, a grid-aligned square is preferable to a circle at the chosen resolution. With ten cells per wavelength, a circle of diameter nine cells is represented by a stepped boundary that can differ from the intended surface by up to half a cell, or $\lambda/20$, and its area differs from the ideal circle by about $8.5\%$. A square, by contrast, is represented exactly on the grid. This avoids introducing a geometric discretisation difference between the two solvers, so that differences in their results can be attributed to the encoder rather than to different representations of the scatterer.}

\RM{The scattering case is also a useful benchmark because the obstacle breaks the near-degenerate mode cluster that is strongly excited in the empty cavity. This reduces the effective quality factor and therefore the resonant build-up that the ternary scheme has difficulty reproducing. With the scatterer present, the probe amplitude ratio increases from $0.56$ to $0.91$. The improved agreement therefore suggests that the large amplitude deficit in the empty cavity is associated primarily with the strong resonant build-up of the undamped cavity, rather than with the ternary representation itself.}

Algorithm~\ref{alg:step} can be implemented in both software and hardware using BitNet-like compact data packing~\cite{Maksymov2026TernaryVHDL}.

\RII{\section*{Data availability}
The source code is available in the repository cited in Ref.~\cite{Maksymov2026TernaryFDTD}.}

\RII{\section*{Conflict of interest}
The author states no conflict of interest.}

\bibliographystyle{eplbib}
\bibliography{epl}

\end{document}